# Optimize the Co-expansion of Generation and Transmission Considering Wind Power in the US Eastern Interconnection

Shutang You

## Abstract

This paper studies the generation and transmission expansion co-optimization problem with a high wind power penetration rate in large-scale power grids. In this paper, generation and transmission expansion co-optimization is modeled as a mixed-integer programming (MIP) problem. A scenario creation method is proposed to capture the variation and correlation of both load and wind power across regions for large-scale power grids. Obtained scenarios that represent load and wind uncertainties can be easily introduced into the MIP problem and then solved to obtain the co-optimized generation and transmission expansion plan. Simulation results show that the proposed planning model and the scenario creation method can improve the expansion result significantly through modeling more detailed information of wind and load variation among regions in the US EI system. The improved expansion plan that combines generation and transmission will aid system planners and policy makers to maximize the social welfare in large-scale power grids.

## Nomenclature

Parameters

| | |
|---|---|
| $CF_{y,r,s}$ | Wind capacity factor of generator $g$ in region $r$, year $y$, scenario $s$ |
| $C_{emm,r,y,g}$ | Emission cost for generator $g$, region $r$, year $y$ |
| $C_{FOM,r,g}$ | Fixed operation and maintenance cost of generator $g$ located in region $r$ |
| $C_{fuel,y,r,g}$ | Fuel Price for generator $g$, region $r$, year $y$ |
| $C_{Gbuilt,y,r,g}$ | Build cost of generator $g$ located in region $r$, year $y$ |
| $C_{VOLL,r}$ | Value of lost load (energy shortage price) in region $r$ |
| $C_{VOM,r,g}$ | Varying operation and maintenance cost of generator $g$ in region $r$ |
| $C_{Wheeling,l}$ | The wheeling cost coefficient in interface $l$ |
| $C_{Xbuilt,y,l}$ | Build cost of transmission interface $l$ in year $y$ |
| $DF_y$ | Discount factor in the year $y$ |
| $e_{r,g}$ | Emission coefficient of generator $g$ located in region $r$ |
| $F_{FOR,r,g}$ | Forced outage rate of generator $g$ located in region $r$ |
| $L'$ | Load duration curves of raw chronological load data |
| $L_{y,r,s}$ | Load in region $r$, scenario $s$, year $y$ |
| $L_{y,s}$ | System-level load in scenario $s$, year $y$ |
| $MF_{r,s}$ | Maintenance factor in region $r$ scenario $s$ |
| $N_L$ | Number of transmission interfaces between regions |
| $N_Q$ | Number of raw chronological curves |
| $N_R$ | Number of regions |
| $N_{r,G}$ | Number of generators (including both existing and candidates) in region $r$ |
| $N_Y$ | Number of years in the planning horizon |
| $N_{y,S}$ | Number of scenarios in year $y$ |
| $P_{\max,r,g}$ | Maximum generation capacity of generator $g$ in region $r$ |
| $\overline{P}_{\max,r,g,s}$ | Maximum generation of generator $g$ after considering the forced and maintenance outages discount in scenario $s$ in region $r$ |
| $P_{l,\max}$ | Maximum transmission capacity of transmission interface $l$ |
| $pr_{y,s}$ | Probability of scenarios $s$ in year $y$ |
| $pr'_q$ | Probability of raw load curves $q$ |
| $R_{H,r,g}$ | Heat rate of generator $g$ in region $r$ |
| $Rs_{y,r}$ | Required reserve margin in region $r$ year $y$ |
| $RPS_{y,r}$ | Renewable portfolio standard percentage in region $r$ year $y$ |
| $T$ | Number of hours in each year |
| $x^0_{r,g}$ | Number of existing units of generator $g$ |
| $\bar{x}_{Gbuilt,y,r,g}$ | Max unit number of annual expansion of generator $g$ in region r |
| $X_{MaxGbuilt,r,g}$ | Maximum number of unit expansion of generation $g$ |
| $X_{MaxXbuilt,l}$ | Maximum number of unit expansion of interface $l$ |
| $\bar{x}_{Xbuilt,y,l}$ | Max unit number of annual expansion of interface $l$ |
| $x^0_{r,l}$ | Number of existing lines of interface $l$ |
| $\Omega_r$ | Set of the transmission interfaces of region $r$ |

Variables

| | |
|---|---|
| $I_{y,l,s}$ | Power flow of transmission interface $l$ in scenario $s$, year $y$ |
| $P_{y,r,g,s}$ | Dispatch level of generating unit $g$ in scenarios s, region $r$, and year $y$. |
| $P_{USE,y,r,s}$ | Unserved power in region $r$, scenario $s$, year $y$ |
| $x_{Gbuilt,y,r,g}$ | Expansion decision variable of generator $g$ |
| $x_{Xbuilt,y,l}$ | Expansion decision variable of transmission interface $l$ in year $y$ |

Indices

| | |
|---|---|
| $g$ | Generator index |
| $l$ | Transmission interface index |
| $q$ | Raw chronological curve index |
| $r$ | Region index |
| $s$ | Scenario index |
| $y$ | Year indexes |

## 1. INTRODUCTION

Bulk power system expansion problems can be divided into three categories: generation expansion [1, 2], transmission expansion [2-4], and generation-transmission co-expansion [5]. Power system operation is subjected to influences from stochastic factors, such as forced outages, load, and fuel cost variations. With the increase of renewable penetration rates, the stochastic features of wind and solar are becoming the major uncertain factors of power systems. As studies predict that US could have around 27% of its electricity coming from renewables by 2030 [6], their fluctuations need to be considered not only in the operation stage through advanced monitoring and control, but also in the planning stage.

### 1.1 Literature Review

Much research focused on expansion planning for generation and transmission. Several methods exist to categorize these studies. For example, studies can be categorized by single-stage [7-9] or multiple-stage formulations of the problem [10-14], which can provide the timing of expansion. Research can also be classified by its solving algorithm: conventional mathematical programming or meta-heuristic optimization methods. While meta-heuristic algorithms possess the capability to deal with non-linear and non-convex optimization problems [15-19], they have the defect of converging to local-optima. Conventional mathematical programming, such as mixed-integer programming, are more computational efficient but their applicability is more restricted by the model characteristics. Some recent research seeks to solve a series of mixed-integer linear programming problems iteratively to obtain the optimal generation expansion plan [1, 20].

Existing studies can also be classified by their problem formulation structures: some research addresses generation or transmission expansion only; while other studies optimize both the generation and the transmission expansion plan. Research that considers both generation and transmission expansion further falls into two sub-categories. One category is the multi-level or hierarchical models that apply iterative or decomposition methods to obtain the expansion result. For example, Ref. [2, 21, 22] formulated the generation transmission expansion plan as tri-level problems that consist of a pool-based market, generation expansion and transmission expansion. The other category is formulating the two expansion problems in one problem (co-optimization). For example, Ref. [8] proposed a mixed integer nonlinear programming formulation for generation and transmission planning considering the lost load cost, and then the problem was linearized to a mixed integer linear programming problem. In [23-25], the expansion planning is directly formulated as mixed integer programming problems with reliability constraints enforced iteratively. As extensions of the previous work based on the MIP formulation, Ref. [26] applied a modified genetic algorithm to solve the MIP problem and adopted a virtual database to accelerate detailed reliability assessment. Ref. [27] proposed a co-optimizing model with multiple objectives related to renewable energy development.

As renewables grow in power grids, the influence of their uncertainties on generation and transmission planning has drawn much attention. Various methods have been applied to consider wind and load uncertainties. These methods include the fuzzy decision approach [28], the chance-constrained model [29], and the benders decomposition [7, 30, 31]. Apart from modelling uncertainty of wind in one spot, [32] further considered the uncertainties of wind speed correlation in integrating wind to a power grid, and the optimal integration plan was obtained using fuzzy techniques. To provide multiple plan candidates, Ref. [33] applied probabilistic power flow to consider the uncertainty of wind power and a genetic algorithm to obtain Pareto optimal solutions of a multi-objective transmission expansion problem.

Recent progresses on robust optimization techniques can consider uncertain parameters in expansion optimization [32, 34, 35]. Two categories of robust optimization models have been successfully applied in power system expansion: probabilistic robust optimization models (e.g. stochastic optimization models) [32, 35] and non-probabilistic robust optimization models [34]. Probabilistic robust optimization models are capable of considering detailed parameter distribution such as wind and load variation. The most commonly-used techniques for solving a stochastic optimization problem is formulating its deterministic equivalent through scenario construction. Besides considering all possible combinations of uncertainties [36, 37], some techniques to select representative scenarios have been applied, such as Taguchi's Orthogonal Array Testing [38]. As the other category of robust optimization, non-probabilistic robust optimization models are more adaptive to the representation of uncertainties. Therefore, they are applicable when parameter probabilistic distribution is not available. Such uncertainties include policy changes and market participant behaviors. Non-probabilistic optimization models usually adopt the Wald's maximum model to minimize of the maximum adjustment cost and regret [34, 39]. The deficiencies of some robust optimization models include the computation complexity when applied to large-scale systems (for probabilistic robust optimization models) and the lack of ability to quantify the overall expected cost (for non-probabilistic robust optimization models). Besides, scenario construction to balance model accuracy and complexity is still an active research topic for both categories of methods [40].

## 1.2 Aim and Contributions

It has been widely accepted that co-optimization generation and transmission expansion can obtain better results and more investment savings [41]. With this combined planning information, policy makers can formulate market rules, incentives and penalties to make sure regional planning authorities and stakeholders are expanding and operating the power grid in a way close to maximizing social welfare. Moreover, the variation of renewables in different regions has significantly increased interface flow and energy exchange between regions, requiring combined generation-transmission expansion.

A major obstacle for implementing existing methods to actual systems is the computation complexity of the Level II problems, which requires multiple year's operation or market simulation. The computation burden further grows exponentially with renewables, fuel price and load uncertainties in the temporal and spatial dimension. Additionally, few studies have investigated co-optimizing generation and transmission expansion considering renewables in large-scale multi-region power grids [42]. This paper presents a mixed-integer programming (MIP) formulation that can co-optimize generation and transmission expansion considering wind power and load variations on a multi-region basis. Contributions of this paper include:

- This paper proposes an uncertainty-incorporated MIP formulation to co-optimize generation and transmission expansion. Under this framework, the timing and capacity of generation and transmission expansion can be obtained through solving a single optimization problem.
- A scenario generation method is proposed to incorporate the demand and wind diversity across regions in a large-scale power grid. Incorporating these diversities into the co-optimization model facilitates more accurate modelling of inter-regional energy exchange and interface expansion.
- Actual system data of the US Eastern Connection (EI) system with a 15-year planning horizon are used to verify the proposed method's capability for implementing in actual large-scale power grids. In addition, the obtained result can aid system planners and policy makers to maximize the social welfare of the EI system.

The structure of this paper is organized as follows. Section 2 presents the formulation of the generation and transmission co-optimization model. Section 3 introduces the scenario generation method. Section 4 describes the dataset, study cases, and results based on the US EI system. Section 5 gives the conclusions.

## 2. CO-OPTIMIZING GENERATION-TRANSMISSION EXPANSION

The generation and transmission expansion model aims to maximize the social welfare or minimize the total cost, which is comprised of the capacity expansion cost, the operation cost, and the emission cost over the planning horizon. The capacity expansion cost includes both the generation and the transmission expansion cost. The operation cost consists of the fixed operation cost and maintenance cost, the varying operation and maintenance cost, the fuel cost of generators, the value of the lost load, and the wheeling cost of transmission lines. Based on a generic MIP model [43], the objective function of the generation and transmission expansion co-optimization problem can be expressed as:

$$f = \sum_{y=1}^{N_Y}\sum_{r=1}^{N_R}\sum_{g=1}^{N_{r,G}} DF_y \cdot \left(C_{Gbuilt,y,r,g} \cdot x_{Gbuilt,y,r,g}\right)$$

$$+ \sum_{y=1}^{N_Y}\sum_{r=1}^{N_R}\sum_{g=1}^{N_{r,G}} DF_y \cdot \left[C_{FOM,r,g} \cdot P_{\max,r,g}\left(x^0_{r,g} + \sum_{y'\le y} x_{Gbuilt,y',r,g}\right)\right]$$

$$+ \sum_{y=1}^{N_Y}\sum_{r=1}^{N_R} DF_y \cdot T \sum_{s=1}^{N_{y,S}} pr_{y,s} \cdot \sum_{g=1}^{N_{r,G}} R_{H,r,g} \cdot C_{fuel,y,r,g} \cdot P_{y,r,g,s}\left(x^0_{r,g} + \sum_{y'\le y} x_{Gbuilt,y',r,g}\right)$$

$$+ \sum_{y=1}^{N_Y}\sum_{r=1}^{N_R} DF_y \cdot T \sum_{s=1}^{N_{y,S}} pr_{y,s} \cdot \sum_{g=1}^{N_{r,G}} C_{VOM,r,g} \cdot P_{y,r,g,s}\left(x^0_{r,g} + \sum_{y'\le y} x_{Gbuilt,y',r,g}\right)$$

$$+ \sum_{y=1}^{N_Y}\sum_{r=1}^{N_R} DF_y \cdot T \sum_{s=1}^{N_{y,S}} pr_{y,s} \cdot C_{VOLL,r} \cdot P_{USE,y,r,s}$$

$$+ \sum_{y=1}^{N_Y}\sum_{l=1}^{N_L} DF_y \cdot \left(C_{Xbuilt,y,l} \cdot x_{Xbuilt,y,l}\right)$$

$$+ \sum_{y=1}^{N_Y}\sum_{l=1}^{N_L} DF_y \cdot T \sum_{s=1}^{N_{y,S}} pr_{y,s} \cdot C_{Wheeling,l} \cdot I_{y,l,s} \cdot \left(x^0_{r,l} + \sum_{y'\le y} x_{Xbuilt,y',l}\right)$$

$$+ \sum_{y=1}^{N_Y}\sum_{r=1}^{N_R}\sum_{g=1}^{N_{r,G}} DF_y \cdot \left[C_{emm,y,r,g} \cdot e_{r,g} \cdot T \sum_{s=1}^{N_{y,S}} pr_{y,s} \cdot P_{y,r,g,s}\left(x^0_{r,g} + \sum_{y'\le y} x_{Gbuilt,y',r,g}\right)\right] \quad (1)$$

where $DF_y$ is the discount factor in year $y$. Before the end of the planning horizon, its value is:

$$DF_y = \frac{1}{(1+d)^y} \qquad y = 1,2,...,N_Y - 1 \quad (2)$$

It is important that the expansion planning formulation does not inappropriately consider the end of the planning horizon to be the 'end of time'. Without considering the 'end-year effects', the expansion plan would select to build generators with low build costs in the last several years, even if their marginal generation costs are high, so that the average cost in the horizon would be low. To reflect the 'end-year effects', the last year of the horizon is repeated an infinite number of times [43]. Therefore, the discount factor in the end year considering the 'end-year effects' is

$$DF_{N_Y} = \frac{1}{(1+d)^{N_Y}} + \frac{1}{(1+d)^{N_Y+1}} \cdot \frac{1}{\left(1 - \frac{1}{1+d}\right)} \quad (3)$$

In (1), $\sum_{y=1}^{N_Y}\sum_{r=1}^{N_R}\sum_{g=1}^{N_{r,G}} DF_y \cdot \left(C_{Gbuilt,r,g} \cdot x_{Gbuilt,y,r,g}\right)$ is the generation built cost in all regions;

$\sum_{y=1}^{N_Y}\sum_{r=1}^{N_R}\sum_{g=1}^{N_{r,G}} DF_y \cdot \left[C_{FOM,r,g} \cdot P_{\max,r,g}\left(x^0_{r,g} + \sum_{y'\le y} x_{Gbuilt,y',r,g}\right)\right]$ is the fixed operation and maintenance cost in all regions (in \$/kW/year, forming part of the unit annual fixed cost charge);

$$\sum_{y=1}^{N_Y}\sum_{r=1}^{N_R} DF_y \cdot T \sum_{s=1}^{N_{y,S}} pr_{y,s} \cdot \sum_{g=1}^{N_{r,G}} R_{H,r,g} \cdot C_{fuel,y,r,g} \cdot P_{y,r,g,s}\left(x_{r,g}^0 + \sum_{y'\leq y} x_{Gbuilt,y',r,g}\right)$$ denotes the fuel cost.

$$\sum_{y=1}^{N_Y}\sum_{r=1}^{N_R} DF_y \cdot T \sum_{s=1}^{N_{y,S}} pr_{y,s} \cdot \sum_{g=1}^{N_{r,G}} C_{VOM,r,g} \cdot P_{y,r,g,s}\left(x_{r,g}^0 + \sum_{y'\leq y} x_{Gbuilt,y',r,g}\right)$$ denotes the varying operation and maintenance cost (in \$/kW/year, representing an incremental cost of generation used to recover regular equipment replacement and servicing costs that are a direct function of generation, e.g. wear and tear);

$$\sum_{y=1}^{N_Y}\sum_{r=1}^{N_R} DF_y \cdot T \sum_{s=1}^{N_{y,S}} pr_{y,s} \cdot C_{VOLL,r} \cdot P_{USE,y,r,s}$$ denotes the value of lost load; $\sum_{y=1}^{N_Y}\sum_{l=1}^{N_L} DF_y \cdot \left(C_{Xbuilt,y,l} \cdot x_{Xbuilt,y,l}\right)$ is the transmission expansion cost of all interfaces;

$$\sum_{y=1}^{N_Y}\sum_{l=1}^{N_L} DF_y \cdot T \sum_{s=1}^{N_{y,S}} pr_{y,s} \cdot C_{Wheeling,l} \cdot I_{y,l,s} \cdot \left(x_{r,l}^0 + \sum_{y'\leq y} x_{Xbuilt,y',l}\right)$$ is the wheeling cost of transmission lines;

$$\sum_{y=1}^{N_Y}\sum_{r=1}^{N_R}\sum_{g=1}^{N_{r,G}} DF_y \cdot \left[C_{emm,y,r,g} \cdot e_{r,g} \cdot T \sum_{s=1}^{N_{y,S}} pr_{y,s} \cdot P_{y,r,g,s}\left(x_{r,g}^0 + \sum_{y'\leq y} x_{Gbuilt,y',r,g}\right)\right]$$ is the emission cost.

The objective function (1) is the net present value of the sum of all of the system's cost items over the planning horizon considering the 'end-year effects'. A practical expansion plan should also satisfy various planning and operation constraints. Constraints considered in this expansion planning problem include:

- Power balance constraint

In each region, the sum of generation output, unserved demand, and interface interchange should equal to the demand for all regions within the planning horizon.

$$\sum_{g=1}^{N_{r,G}} P_{y,r,g,s} + P_{USE,y,r,s} + \sum_{l\in\Omega_r} I_{y,l,s} = L_{y,r,s} \qquad \forall y,\forall r \qquad (4)$$

where $\sum_{g=1}^{N_{r,G}} P_{y,r,g,s}$ denotes generation of all generators in region $r$ under the scenario $s$; $\sum_{l\in\Omega_r} I_{y,l,s}$ denotes the power interchange through all the interfaces of region $r$.

- Maximum expansion constraint for generation

Due to resource limitation (such as maximum exploitable resources for hydro and nuclear power plants), the number of generator expansion in each region should be within its upper limit.

$$\sum_{y\leq N_Y} x_{Gbuilt,y,r,g} \leq X_{MaxGbuilt,r,g} \qquad \forall g,\forall r \qquad (5)$$

- Maximum expansion constraint for transmission

Due to the right-of-way limitation, the number of expanded interfaces should be within its upper limit.

$$\sum_{y \leq N_Y} x_{Xbuilt,y,l} \leq X_{MaxXbuilt,l} \qquad \forall l \qquad (6)$$

- Integer constraint

The number of built generators and interfaces should be integers.

$$x_{Gbuilt,y,r,g} \in N \ ; \ x_{Xbuilt,y,l} \in N \qquad (7)$$

- Expansion speed constraint

Due to the construction resource limitation, the annual expansion speed of generators and transmission lines should be within their upper limits.

$$x_{Gbuilt,y,r,g} \leq \overline{x}_{Gbuilt,y,r,g} ; \ x_{Xbuilt,y,l} \leq \overline{x}_{Xbuilt,y,l} \qquad \forall y, \forall g, \forall r, \forall l \qquad (8)$$

- Capacity discount considering the forced and maintenance outages

In each region, the available capacity is usually less than the installed capacity due to forced and maintenance outages. The capacity discount is determined by the maintenance outage rate, the maintenance factor, and the forced outage rate. Since maintenance tasks are usually scheduled in off-peak periods, the maintenance factor $MF_{r,s}$ is related to the load level in scenario $s$.

$$\overline{P}_{\max,r,g,s} \leq \left(1 - F_{MOR,r,g} \cdot MF_{r,s} + F_{FOR,r,g}\right) \cdot P_{\max,r,g} \cdot \left( x^0_{r,g} + \sum_{y' \leq y} x_{Gbuilt,y',r,g} \right) \qquad \forall g, \forall r \qquad (9)$$

- Regional reserve capacity constraint

The reserve capacity of each region should be larger than a pre-determined level for regulation and contingency requirements. The reserve capacity level is set to be able to cover the energy imbalance caused by any single generator or interface failure associated with this region.

$$\overline{P}_{\max,r,g,s} \left( x^0_{r,g} + \sum_{y' \leq y} x_{Gbuilt,y',r,g} \right) \geq L_{y,r,s} - p \sum_{l \in \Omega_r} I_{y,l,s} + Rs_{y,r} \qquad \forall y, \forall r \qquad (10)$$

- Interface capacity constraint

To facilitate interregional analysis based on information provided by regional planning authorities [44], resources inside a region are considered to be connected to a notional node. Nodes that are not associated with a region is considered as regions by themselves. The interface power flows between regions are modelled by the transportation model [45, 46]. Interface flows should satisfy their limits.

$$-P_{l,\max}\left(x^0_{r,l} + \sum_{y'\leq y} x_{Xbuilt,y',l}\right) \leq I_{y,l,s} \leq P_{l,\max}\left(x^0_{r,l} + \sum_{y'\leq y} x_{Xbuilt,y',l}\right) \quad \forall y, \forall l \quad (11)$$

- Regional renewable portfolio constraint

In those regions with renewable portfolio constraints, the percentage of renewables in the total installed generation capacity should be higher than a pre-determined value.

$$\sum_{g=1}^{N_{r,G}} P_{\max\ r,g}\left(x^0_{r,g\ (renew)} + \sum_{i\leq t} x_{Gbuilt,y,r,g\ (renew)}\right) \geq RPS_{y,r} \cdot \sum_{g=1}^{N_{r,G}} P_{\max\ r,g}\left(x^0_{r,g} + \sum_{y'\leq y} x_{Gbuilt,y',r,g}\right) \quad \forall y, \forall r \quad (12)$$

- Wind resource constraint

The output of wind turbine generators is restricted by the available wind resource.

$$P_{y,r,g,s\ (wind)} \leq CF_{y,r,s} \overline{P}_{\max,\ r,g\ (wind)} \quad \forall y, \forall r, \forall g \quad (13)$$

The objective function (1) and constraints (4) - (13) comprise a mixed-integer programming problem that minimizes the cost of the expansion plan while satisfying multiple operational and environmental constraints.

Except for the fixed operation and maintenance cost, all the operation-related cost items in the objective function are related to the load and wind levels. The obtained expansion plan is the optimal considering all the stochastic factors, such as load and wind. Section 3 will show the methodology of integrating these stochastic factors into the generation and transmission expansion co-optimization problem.

## 3. SCENARIO CREATION FOR MULTI-REGION SYSTEMS

In long-term planning, hourly load data are usually aggregated and represented by blocks (load scenarios) for computation complexity consideration. Typical load aggregation methods include the load duration curve method and the load curve fitting method. The resulting load representation will have a limited number of scenarios in each year that represent different levels of demand and corresponding time duration.

For multi-region expansion planning, load scenarios should be synchronized (the load data in one scenario have the same chronological time stamps for all regions) in order to preserve the load diversity across regions. Load scenario synchronization is accomplished through the following steps:

*Load scenario synchronization for multi-region power grids*

1) Obtain the load duration curves $\{L'\}_{N_Q}$ of the system based on the future demand data obtained by load forecasting. The total number of the curves is $N_Q$ and each curve has probability $pr'_q$.

2) The target scenario load scenarios $\{L_{y,s}\}_{N_{y,s}}$ and probabilities $pr_{y,s}$ are simultaneously optimized using the least square fitting approach, i.e. minimizing $\sum_{s=1}^{N_{y,s}} pr_{y,s} \left( \sum_{q=1}^{N_Q} pr'_q (L_{y,s} - L')^2 \right)$. (In particular, the maximum demand period of all load duration curves are preserved to form a scenario.) Recording the chronology-to-scenario mapping function $\Theta_L$.

3) Use $\Theta_L$ to re-arrange the load data of each region.

4) Calculate the expected value of load data for each scenario and region. The scenario probabilities are inherited from Step 2).

Similar to load, the wind power variation in each region should also be considered in expansion planning, especially if wind generation is a significant fraction of the total generation. In addition, since large power systems usually have wide geographic areas, wind resources could vary greatly in different regions at the same time point. Therefore, it is needed to develop a sub-scenario creation method that can most effectively capture the information of multi-regional chronological wind output. The expansion model can then include multiple scenarios of wind power output in each region, as well as the correlation information of wind power output across regions in each demand level.

To meet this requirement, a generic scenario creation method for multiple stochastic variables in expansion planning is shown in Figure 1. Description of some steps is as follows.

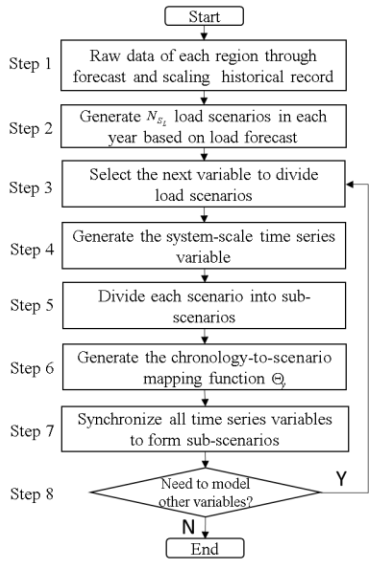

Figure 1. The sub-scenario creation procedure

- Step 1. Prepare the time series load, wind, or solar data over the planning horizon based on forecast data or scaling historical records.
- Step 2. Use the least square fitting method to form $N_{S_L}$ load scenarios, namely $s_{L1}, s_{L2}, ..., s_{LN_{SL}}$, in each year. That is $N_{S_L} \times N_Y$ scenarios in total over the planning horizon. Probability of each scenario is denoted by $pr_{y,s}$.
- Step 3. Select the next stochastic factor $v_k$ ($v_k$ could be wind, solar, or cost uncertainty depending on the influences on expansion. In the following steps, wind is described for convenience).
- Step 4. Create the system-scale time series of the wind capacity factor. In many large-scale power grids, wind farms (both existing and candidate) are centralized in certain regions with rich wind resource, the time-series wind capacity factors in these regions are better indicators of the wind output of the whole system compared with those of the regions that have much less wind power. Thus, the weighted time series value of the wind capacity factors is used to form a system-scale chronological attribute— the multi-region system wind capacity factor $CF_{system,t}$.

$$CF_{system,t} = \sum_{r=1}^{N_R} WindCap_r^* \cdot CF_{r,t} \qquad (14)$$

where $WindCap_r^*$ denotes the rough target wind generation capacity in the region $r$, which can be obtained through estimation or iteratively updating from the previous planning result. $CF_{r,t}$ is the time-series wind capacity factor in region $r$. $N_R$ is the number of regions.
- Step 5. Divide each load scenario $s_{Ll}$ into $N_{S_W}$ sub-scenarios of different wind scenarios, namely $s_{Ll,W1}, s_{Ll,W2}, ..., s_{Ll,WN_{SW}}$. Each sub-scenario represents a certain range of those wind capacity factors happening during the corresponding period of the load scenario. For instance, the lowest $1/N_{S_W}$ percent system-scale wind capacity factor chronological data ($CF_{system,t}$) in the load scenario $s_{Ll}$ is aggregated in sub-scenario $s_{Ll,W1}$, and the highest $1/N_{S_W}$ percent data is aggregated in $s_{Ll,WN_{SW}}$. The probability of each sub-scenario is $pr_{y,s}/N_{S_W}$. (An alternative and more computation-intensive method to obtain the sub-scenarios and probabilities using the least square approach as described before.)
- Step 6. Generate the chronology-to-scenario mapping function $\Theta_y$. In forming each sub-scenario, the correspondence between the original chronological data $CF_{system,t}$ and each sub-scenario $s_{Ll,Ww}$ is recorded as the chronology-to-scenario mapping function denoted by $\Theta$. Note that $\Theta$ differs from year to year and it is denoted by $\Theta_y$ for year $y$.
- Step 7. According to the chronology-to-scenario mapping function $\Theta_y$, synchronize all the chronological data (load, wind, and solar) of all regions to form $(N_{S_L} \times N_{S_W}) \times N_Y = N_S \times N_Y$ scenarios.
- Step 8. Check if there are other stochastic factors (such as solar in high solar systems) that need to form sub-scenarios in expansion planning. If so, return to Step 3. Otherwise, we end the scenario generation process.

This scenario forming method can model multiple coupled stochastic variables for long-term expansion planning through creating scenarios that have various probabilities in a computationally effective way. Moreover, the formed scenarios can be directly put into the mixed-integer programming model, and then solved through commercial mathematical programming solvers to obtain the optimal generation-transmission co-expansion plan. It should be noted that Step 7 synchronizes all the stochastic factors, so those factors that have not been selected to further divide scenarios still have different values in different scenarios. The difference is that their scenarios are passively formed (or synchronized) since their information is not specifically used to form sub-scenarios due to that the computation complexity restricts the number of scenarios in the optimization problem.

## 4. IMPLEMENTATION ON THE US EI SYSTEM

### 4.1 Dataset and case description

The proposed model and the scenario generation method is applied to the United States Eastern Interconnection (EI) multi-regional dataset from Charles River Associates [44]. This dataset partitions the EI system into 25 regions and the interfaces between adjacent regions as shown in Figure 2 [47]. The planning horizon is from 2015 to 2030. Five developed cases with different scenario formation methods and number of scenarios are shown in Table 1. The EI multi-regional dataset and the generation and transmission expansion problem are modelled in PLEXOS [43] and the problem is solved by the MIP solver —Xpress-MIP 26.01.04. The optimization is performed on a server with two Intel Xeon E5-2470.0 V2 CPUs (2.40 GHz, 40 cores) and 128 GB memory.

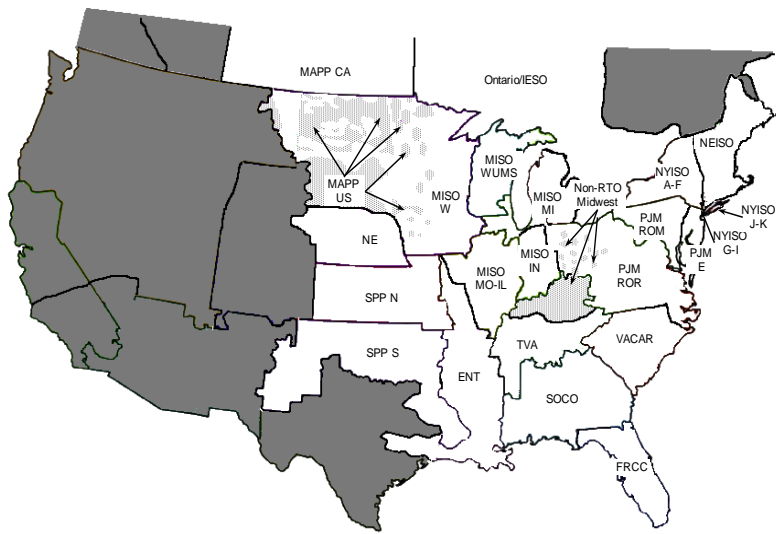

Figure 2. Regions of the U.S. EI system (EI includes all regions in the east) [48]

Table 1. Description on the developed cases

| Case Name | Number of sub-scenarios | Case Description |
|---|---|---|
| 20 Scenarios (20-Scn) | $N_{S_w} = 1$ | • The base case has 20 load scenarios in each year<br>• Wind is the average value in each load scenario |
| 40 Scenarios Non-Synchronized (40-Scn-NonSync) | $N_{S_w} = 2$ | • Splitting each scenario in two equal number of hours<br>• Average of high wind in a half and average of low wind in the other half (*wind data are not synchronized*). |
| 40 Scenarios Synchronized (40-Scn-Sync) | $N_{S_w} = 2$ | • Determining hours of high and low wind capacity factors based on the weighted average system-scale data in each of the 20 load scenarios<br>• Synchronizing wind, solar, and load to the average of the region's values in those hours |
| 80 Scenarios Synchronized (80-Scn-Sync) | $N_{S_w} = 4$ | • Breaking each load scenario into four quartiles based on the weighted average system-scale data in each of the 20 load scenarios<br>• Synchronizing all regions' wind, solar, load, and fuel prices to those hours |
| 160 Scenarios Synchronized (160-Scn-Sync) | $N_{S_w} = 8$ | • Breaking each load scenario into eight quartiles based on the weighted average system-scale data in each of the 20 load scenarios<br>• Synchronizing all regions' wind, solar, load, and fuel prices to those hours |

### 4.2 Comparison and analysis on the expansion results

Applying the co-optimization model, the expansion results of the five cases are summarized in Table 2. It can be noted that the planning results of Case 40-Scn-Sync is between that of Case 20-Scn and Case 40-Scn-NonSync. Since Case 20-Scn only includes one wind output scenario (i.e. the average wind output) in each load scenario, it overestimates the capacity of wind power and underbuilds transmission capacity. Compared with 40-Scn-Sync, the 40-Scn-NonSync Case underestimates the capacity value of wind power since it assumes all regions' wind power is at the high or low half simultaneously, which also leads to more transmission expansion. The 80-Scn-Sync and 160-Scn-Sync Cases add less wind than 40-Scn-Sync but more transmission. This is because the two cases modelled higher wind peak generation scenarios, which need more transmission capacity to export. In the meantime, modelling lower wind scenarios reduces wind power's capacity value, thereby reducing wind power expansion in the planning result.

Table 2. The expansion result summary of the five cases

| Expansion results | 20-Scn | 40-Scn-NonSync | 40-Scn-Sync | 80-Scn-Sync | 160-Scn-Sync |
|---|---|---|---|---|---|
| Wind Candidates Built Capacity[a] (GW) | 262 | 218 | 223 | 221 | 218 |
| All Gen Built Capacity (GW) | 407 | 373 | 381 | 380 | 378 |
| Wind Capacity in 2030 (GW) | 304 | 260 | 265 | 263 | 260 |
| Wind Generation in 2030 (TWh) | 917 | 768 | 783 | 776 | 766 |
| All Gen Build Cost (NPV) (billion $) | 649 | 595 | 603 | 601 | 598 |
| Trans Build Cost (NPV) (billion $) | 20.2 | 26.1 | 22.1 | 22.5 | 25.0 |
| Emission in 2030 (million ton) | 305 | 365 | 358 | 362 | 368 |
| Fuel Offtake 2030 (million GBTU) | 17.1 | 18.2 | 18.1 | 18.1 | 18.2 |

[a] Excluding wind power that has already been decided to build.

The transmission expansion over the planning horizon of the five cases is shown in Figure 3. The number of expanded interfaces of Case 20-Scn is smaller than that of Case 160-Scn-Sync. It indicates that using more detailed wind scenarios will make the transmission expansion more dispersed in space. Particularly, compared with 20-Scn, both 40-Scn-Sync, 80-Scn-Sync, and 160-Scn-Sync have smaller transmission expansion on the interface MISO_IN to PJM_ROR.

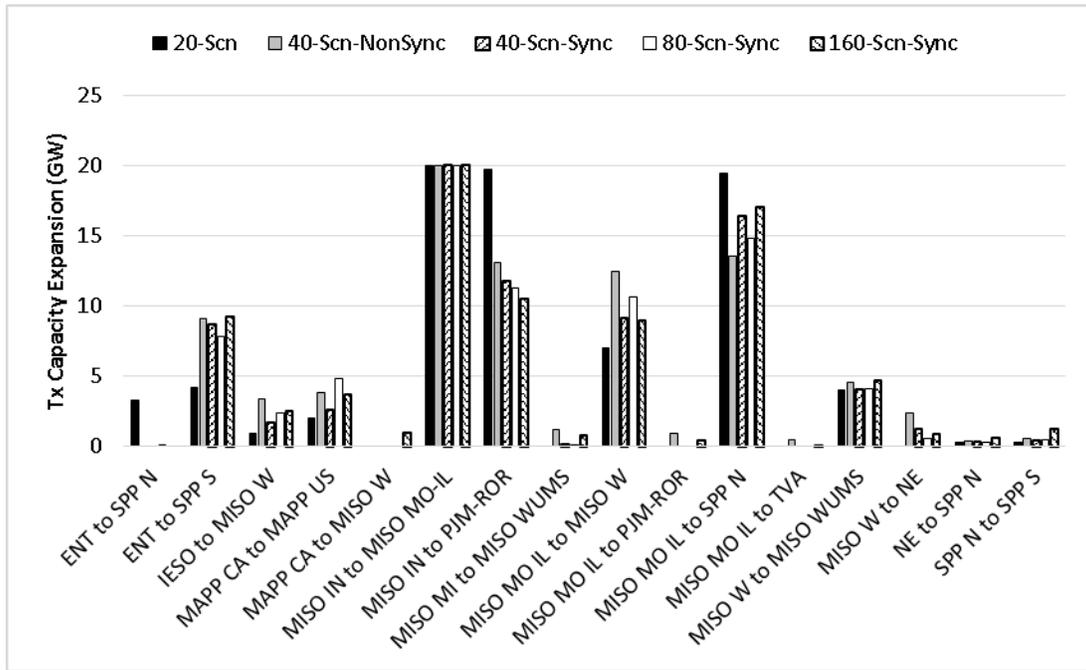

Figure 3. Transmission expansion over the planning horizon in the five cases (Refer to Fig.2 for the labels)

Table 3 shows the expansion of natural gas and wind generation capacity in PJM_ROR and SPP_N. Figure 4 shows the energy flow in 2030 for the Case 20-Scn and Case160-Scn-Sync. Combining Table 3 and Figure 4, it can be seen that in the 20-Scn Case, a large proportion of import energy to PJM_ROR comes from wind in SPP_N. When detailed wind scenarios are incorporated (such as in Case 160-Scn-Sync), PJM_ROR relies more on its local natural gas plants.

Table 3. Expansion of gas and wind generation in Region PJM_ROR and SPP_N

| Expansion Results | 20-Scn | 40-Scn_NonSync | 40-Scn-Sync | 80-Scn-Sync | 160-Scn-Sync |
|---|---|---|---|---|---|
| PJM_ROR Gas Combined Cycle Built (GW) | 6 | 12 | 15.5 | 16 | 17.5 |
| SPP_N Wind Built[a] (GW) | 76.8 | 37.4 | 41.0 | 37.4 | 37.4 |
| PJM_ROR Net Interchange | 153 TWh Import | 82 TWh Import | 78 TWh Import | 69 TWh Import | 61 TWh Import |

[a] Excluding wind power that has already been decided to build.

In addition, it can be noted from Figure 4 that the annual energy flow of almost all interfaces in 160-Scn-Sync increase except for those on the major wind power delivery corridor: SPP_N – MISO_MO_IL – MISO_IN – PJM_ROR. This indicates that detailed wind scenarios will increase the energy exchange frequency and amount between adjacent regions, while decreasing the economy of enforcing transmission networks to transmit a large amount of wind power through a long distance.

For comparison, Figure 5 shows the annual energy flow of the not-co-optimized case (which optimizes transmission and generation expansion sequentially [49]). It can be seen that the not-co-optimized solution chooses to expand the interface between MISO_W and PJM_ROR. In fact, expansion of this interface requires high investment, making the whole expansion plan uneconomic. On the contrary, the co-optimized case is able to leverage multiple regions with good wind resources (e.g. MISO_W and SPP_N) and transfer this energy through an optimal path to regions with high load and low wind (e.g. PJM).

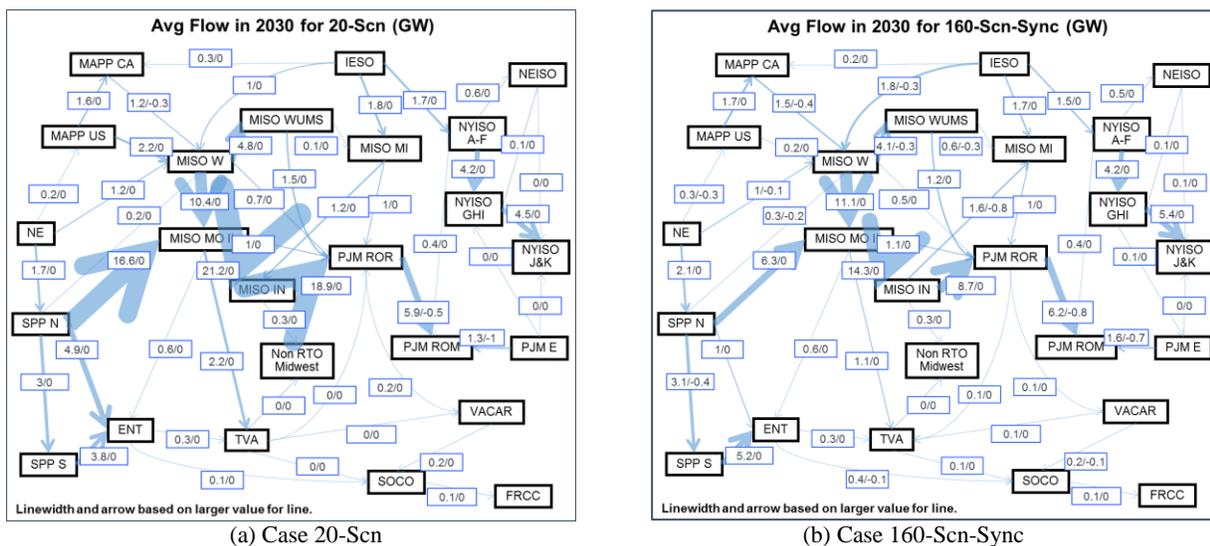

(a) Case 20-Scn  (b) Case 160-Scn-Sync

Figure 4. Annual energy flow of Case 20-Scn and 160-Scn-Sync (Linewidth is proportional to interface annual energy flow)

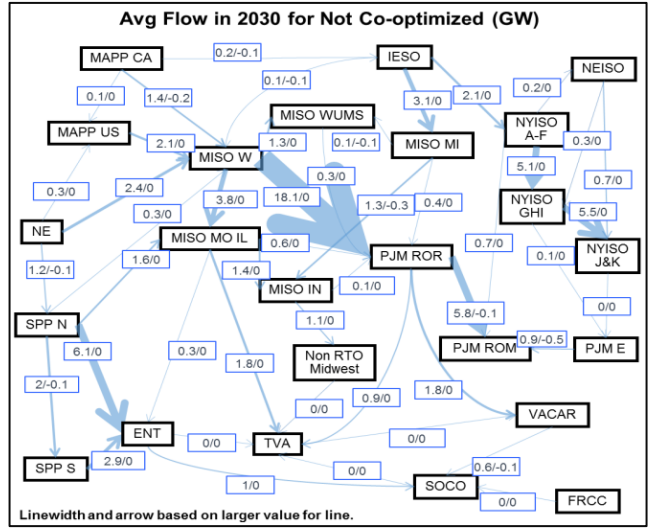

Figure 5. Annual energy flow of the not-co-optimized case (Linewidth is proportional to interface annual energy flow)

### 4.3 Long-term (LT) and short-term (ST) simulation results comparison

In order to quantify the accuracy improvement through the proposed scenario generation method, the long-term expansion result is compared with the short-term simulation result for each case. The long-term simulation applies economic dispatch based on the scenarios generated in the expansion planning model, while short-term simulation uses unit commitment and economic dispatch based on the chronological hourly data. The LT-and ST comparison result is shown in Table 4. It can be noted that there are always gaps between the short-term and long-term results. This is because long-term expansion uses the aggregated scenarios that omit some information in the hourly data. In addition, it shows that Case 160-Scn-Sync has the smallest difference between long-term and short-term simulations, indicating that the operation simulation in Case 160-Scn-Sync is closest to short-term realistic operation. Therefore, on the basis of more accurate modelling of the system operation, the expansion co-optimization result obtained in Case 160-Scn-Sync is more reasonable (but at the expense of execution time.)

Table 4. LT and ST simulation results in 2030

| Results | LT/ST | 20-Scn | 40-Scn-NonSync | 40-Scn-Sync | 80-Scn-Sync | 160-Scn-Sync |
|---|---|---|---|---|---|---|
| Generation cost (NPV billion $) | LT | 47.9 | 55.0 | 54.4 | 54.9 | 55.2 |
|  | ST | 60.3 | 61.9 | 60.0 | 59.7 | 59.5 |
| Emission cost (NPV billion $) | LT | 42.8 | 51.0 | 50.0 | 50.5 | 51.3 |
|  | ST | 50.9 | 52.6 | 52.5 | 53.1 | 53.2 |
| LT-ST gap |  | 18.36% | 7.48% | 7.13% | 6.57% | 5.52% |
| Computation time |  | 1h 6min | 1h 14min | 1h 28min | 4h 28min | 12h 30min |

## 5. CONCLUSIONS

In this paper, generation and transmission expansion is co-optimized using the proposed MIP model, which can be solved robustly using MIP solvers to obtain the global optimal solution. A scenario creation method is proposed to represent wind and load diversities of different regions effectively, thus modelling the interchange of energy between regions more accurately. This scenario creation method can efficiently incorporate uncertainties and operation details into the MIP model for obtaining better expansion plan. The co-optimization model and the scenario creation method are verified through comparing the long-term and short-term simulation results of the US EI system. Additional findings in implementing the proposed framework to the case study are:

(1) Compared with separated optimization, the co-optimized model is able to better leverage wind resources and find a cost-effective path to transmit energy to regions with high load and low wind. The MIP formulation features a systematic consideration of generation and transmission expansion resources.

(2) Incorporating the diversity of wind speed through using more scenarios will decrease the wind expansion capacity and make transmission expansion more dispersed in space. In addition, detailed wind scenarios will reveal that it may be less economic to expand transmission networks to transmit a large amount of wind power through a long distance in the EI system.